# *Frequency Tracking: LMS and RLS Applied to Speech Formant Estimation*

## 1) Introduction

Several speech processing algorithms assume the signal is stationary during short intervals (approximately 20 to 30 ms). This assumption is valid for several applications, but it is too restrictive in some contexts. This work investigates the application of adaptive signal processing to the problem of estimating the formant frequencies of speech. Two algorithms were implemented and tested. The first one is the conventional Least-Mean-Square (LMS) algorithm, and the second is the conventional Recursive Least-Squares (RLS) algorithm.

The formant frequencies are the resonant frequencies of the vocal tract. The speech is the result of the convolution between the excitation and the vocal tract impulse response [Rabiner, 78], thus a kind of "deconvolution" is required to recover the formants. This is not an easy problem because one does not have the excitation signal available.

There are several algorithms for formant estimation [Rabiner, 78], [Snell, 93], [Laprie, 94]. One popular scheme is to calculate an all-pole filter H(z) using the Linear Predictive Coding (LPC) technique based on Levinson-Durbin algorithm. The formants can be associated to the poles of H(z).

The formants can be estimated through an adaptive algorithm as shown in [Hodgkiss, 81]. This work will present results for the conventional LMS and RLS algorithms.

## 2) Discussion on the implemented adaptive algorithms

It is well know that the LMS achieves a very good convergence behavior when the eigenvalues of the signal autocorrelation matrix R are similar. In this case, the negative gradient indicates directly the Wiener solution in the error surface. However, when there is a considerable spread in the eigenvalues, the performance of the LMS algorithm considerably deteriorates.

When the eigenvalues of the correlation matrix R are widely spread, the excess mean-squared error produced by the LMS algorithm is primarily determined by the largest eigenvalues, and the time taken by the average tap-weight vector to converge is limited by the smallest eigenvalues.

Table 2.1 shows the autocorrelation $r(i)$ and eigenvalues $v(i)$ for the speech signal $x$ corresponding to the phrase "We were away". It can be seen that the eigenvalue spread is significant. This was the initial motivation factor for testing the RLS as an alternative to the LMS algorithm.

Table 2.1 - Sample autocorrelation function $r(i)$ and eigenvalues $v(i)$ for "We were away".

| | $i = 0$ | $i = 1$ | $i = 2$ | $i = 3$ | $i = 4$ | $i = 5$ | $i = 6$ | $i = 7$ | $i = 8$ | $i = 9$ | $i = 10$ |
|---|---|---|---|---|---|---|---|---|---|---|---|
| $r(i)$ | 360.54 | 268.05 | 137.39 | 68.63 | 11.05 | -57.60 | -78.29 | -89.97 | -146.76 | -177.18 | -149.36 |
| $v(i)$ | 1,428.2 | 1,136.3 | 433.6 | 380.3 | 187.6 | 152.8 | 152.1 | 55.7 | 14.1 | 12.8 | 12.4 |

Least-square algorithms aim at the minimization of the sum of the squares of the difference between the desired signal and the model filter output. When new samples of the incoming signals are received at each iteration, the solution for the least-squares problem can be computed in recursive form resulting in the recursive least-squares (RLS) algorithms.

An important feature of the RLS algorithm is that it utilizes information contained in the input data, extending back to the instant of time when the algorithm is initiated. The resulting rate of convergence is therefore typically an order of magnitude faster than the simple LMS algorithm. This improvement in performance, however, is achieved at the expense of a large increase in computational complexity and some stability problems.

For the RLS one minimizes the cost function ($\lambda = 1$ corresponds to the ordinary method of least squares [Haykin, 96, page 564])

$$\varepsilon(n) = \sum_{i=1}^{n} \lambda^{n-1} \left| e(i) \right|^2 \tag{2.1}$$

where $\lambda$ is the *exponential weighting factor* or *forgetting factor*.

The RLS algorithms are known to pursue fast convergence even when the eigenvalue spread of the input signal correlation matrix is large. These algorithms have excellent performance when working in time-varying environments [Diniz, 97].

Because speech is highly non stationary, this is another motivation for investigating the performance of the conventional RLS algorithm in the formant estimation task.

The conventional RLS algorithm was implemented according to [Diniz, 97] and [Haykin, 96]. The theoretical development is well described in these two references and will not be repeated here. The implementation of the RLS algorithm in a Matlab function is shown in the Appendix.

# 3) Results

Firstly the algorithms were compared using artificial signals in order to develop intuition and test the implementations. Some of these preliminary results will be presented in subsection 3.1. The second stage, presented in subsection 3.2, was to test the algorithms with the speech data, comparing them to the results obtained with the Levinson-Durbin algorithm.

In this work the adopted adaptive filtering structure was the conventional structure for adaptive spectral estimation, as investigated in the homework.

## 3.1- Artificial signals

This subsection presents the results for a sinusoidal signal with radian frequency equal to $\pi / 9$. The adaptive filters were designed with just two weights to facilitate geometrical visualization. In this example, the two eigenvalues of the signal correlation matrix were 1.93 and 0.06, that corresponds to an eigenvalue spread of $1.93 / 0.06 \cong 32$. The error surface associated to this example is shown in Figure 3.1. It can be seen that this surface is very different from the one in homework 5, where the eigenvalues had the same value.

The Wiener solution for this example is $W^* = [w_0, w_1]^T = [1.7879, -0.9099]^T$.

Several simulations were conducted. This section shows the results when the LMS algorithm was configured with $\alpha = 0.5$ and the RLS algorithm was configured with the forgetting factor $\lambda = 0.8$. It should be noticed that because this simulation did not involve noise, one could choose a large value for $\alpha$ and a small value for $\lambda$. Indeed, RLS with $\lambda = 0.5$ achieved better results than $\lambda = 0.9$, as expected in this case. Figure 3.2 and 3.3 show the instantaneous error for the LMS and RLS algorithms, respectively. It can be seen that even using $\alpha = 0.5$ and a relatively large value for $\lambda$ ($\lambda = 0.8$), the RLS algorithm converged much faster.

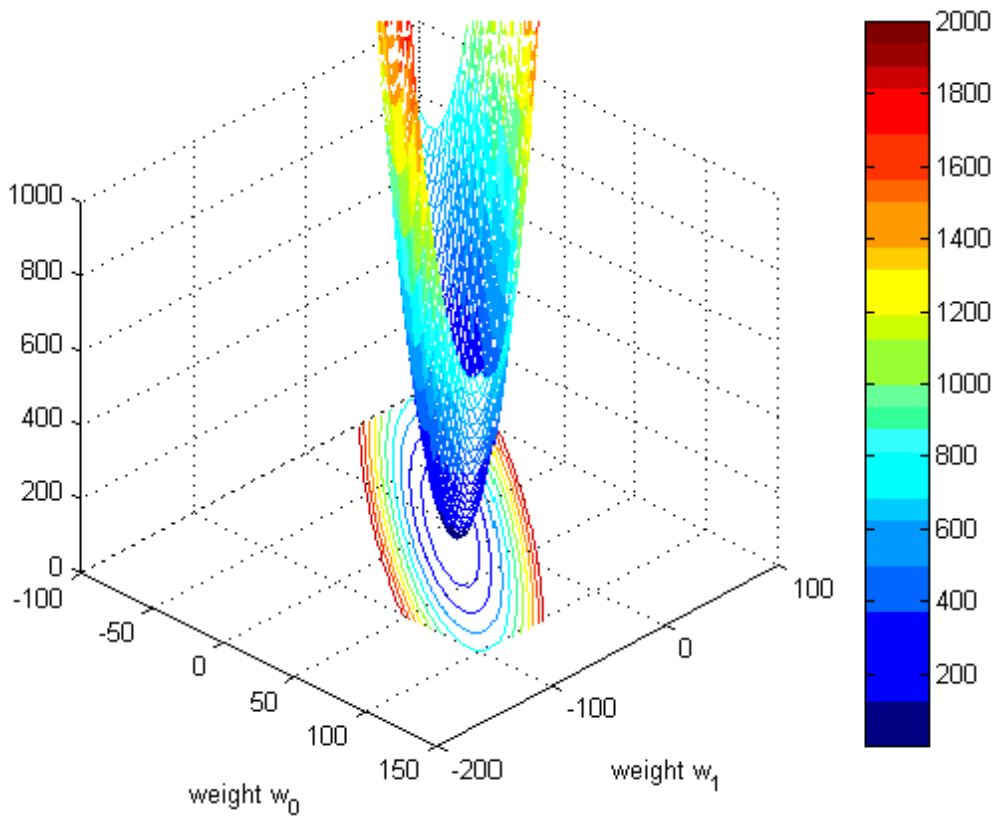

Figure 3.1- Error surface for the case of a signal cos ($\pi$ / 9 n).

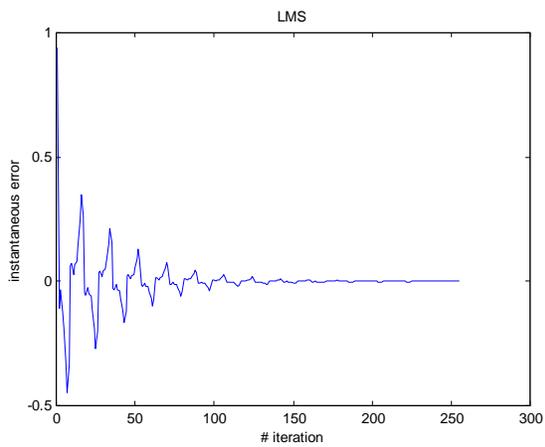

Figure 3.2- Instantaneous error for the LMS algorithm.

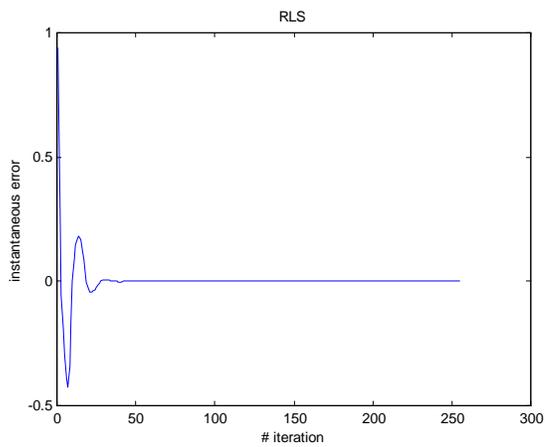

Figure 3.3- Instantaneous error for the RLS algorithm.

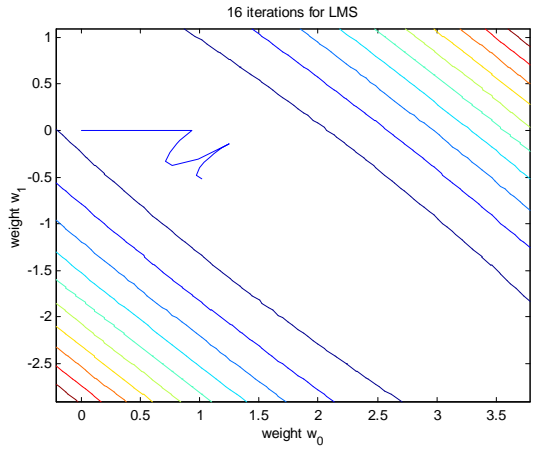

Figure 3.4- Snapshot of LMS convergence.

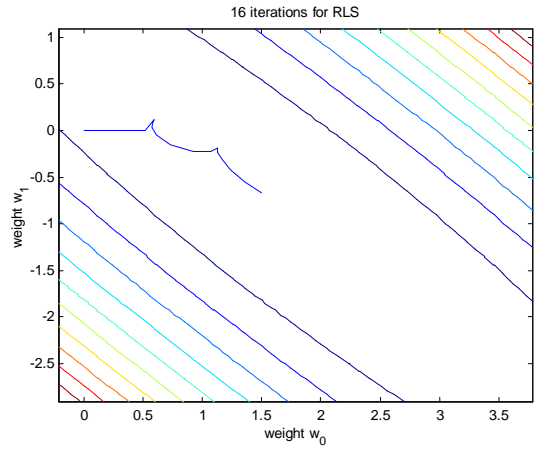

Figure 3.7- Snapshot of RLS convergence.

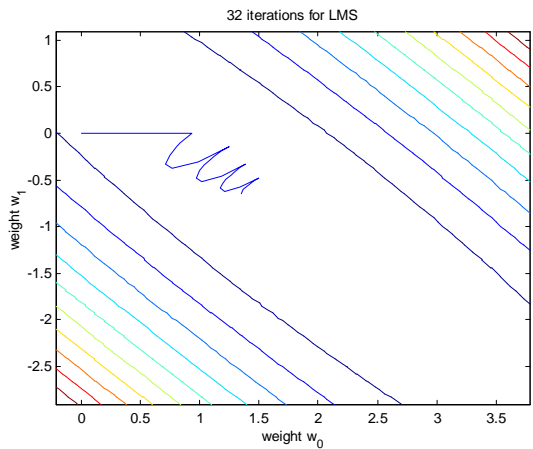

Figure 3.5- Snapshot of LMS convergence.

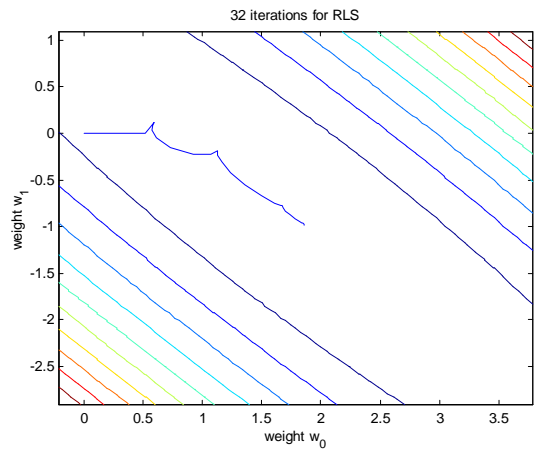

Figure 3.8- Snapshot of RLS convergence.

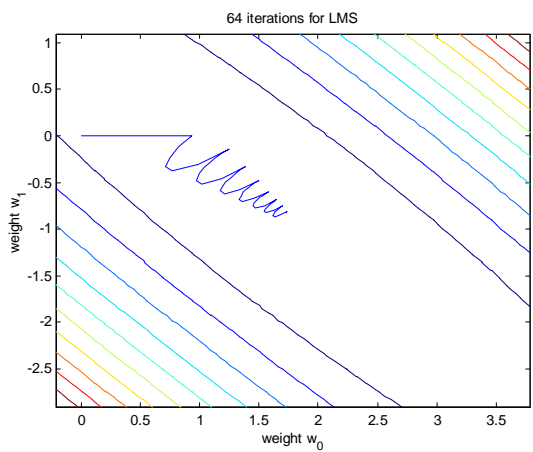

Figure 3.6- Snapshot of LMS convergence.

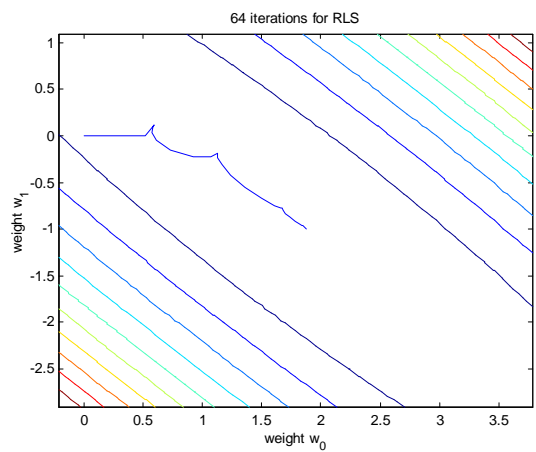

Figure 3.9- Snapshot of RLS convergence.

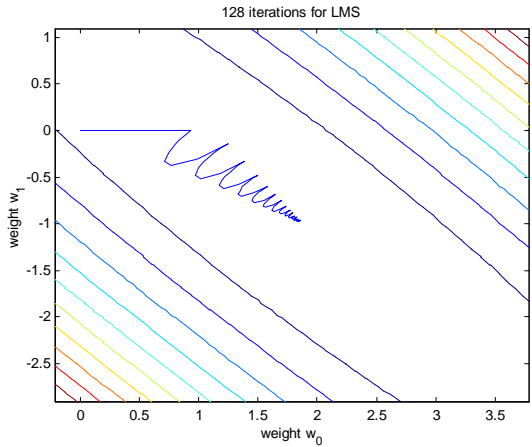

Figure 3.10- Snapshot of LMS convergence.

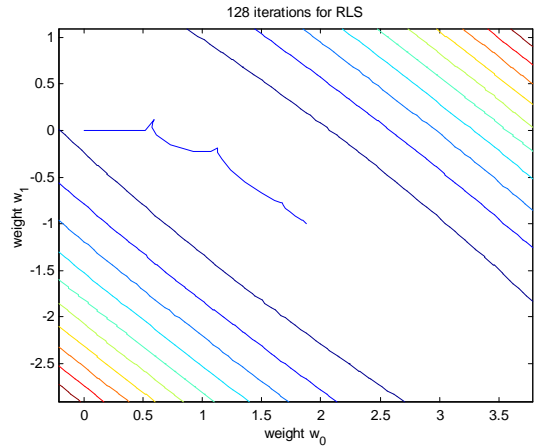

Figure 3.11- Snapshot of RLS convergence.

### *3.2- Speech signal*

This works uses the speech data correspondent to the phrase "We were away", that was digitalized with sampling frequency equal to 8 kHz. The DC value of this signal $x$ was subtracted in all simulations. Figure 3.12 shows a broadband spectrogram (good time resolution and poor frequency resolution) of $x$, calculated through the function *specgram* of Matlab according to the command:

```
specgram(x,1024,8000,blackman(64),round(3/4*64)));
```

that corresponds to having each frame of 64 samples multiplied by a *blackman* window, the frames had an overlap of 3/4 of the frame size, and the spectrum of each windowed frame is calculated through a 1024-points FFT. The zero-padding was used in order to improve the FFT resolution.

In order to give an alternative view, Figure 3.13 shows a narrowband spectrogram (poor time resolution and good frequency resolution) of $x$, calculated according to the command:

```
specgram(x,1024,8000,blackman(256),round(3/4*256)));
```

Figure 3.13 allows one to see the structure due to the pitch (excitation), corresponding to horizontal strips in the plot, while Figure 3.12 shows the pitch influence as vertical strips. For voiced sounds, it is reasonable to assume that the vocal tract is excited just on frequencies that are multiple of the pitch frequency, as Figure 3.13 suggests. If a formant is not localized at a multiple of the pitch, its detection will be more difficult than in the case the formant frequency coincides with a multiple of the pitch.

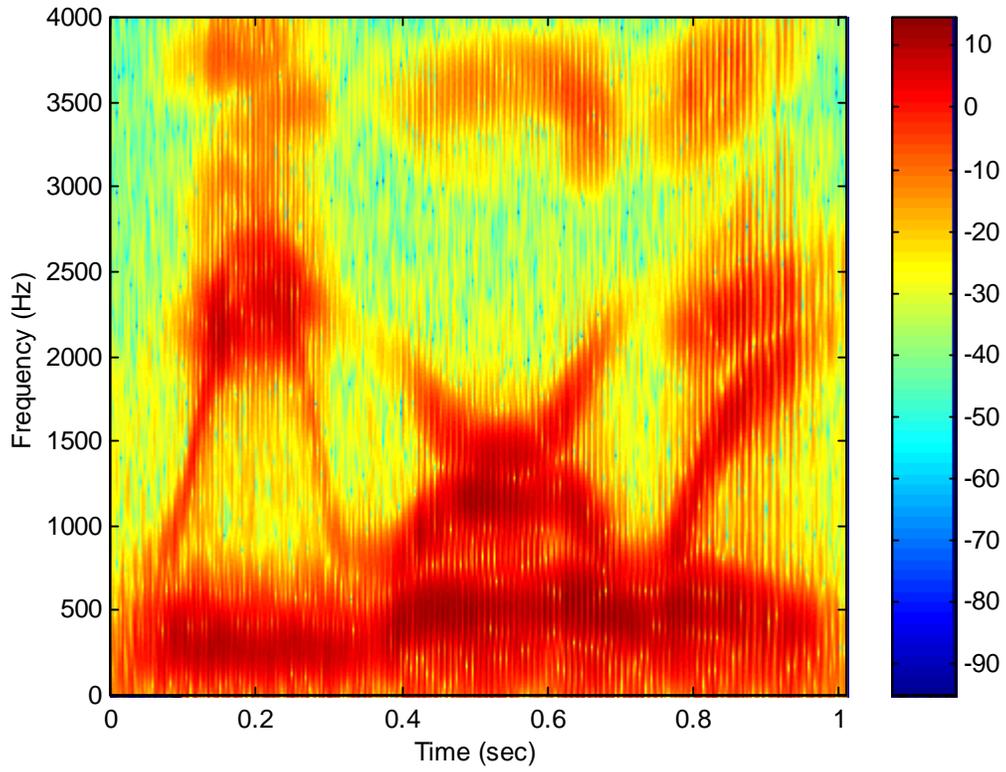

Figure 3.12- Broadband Spectrogram for the phrase "*We were away*".

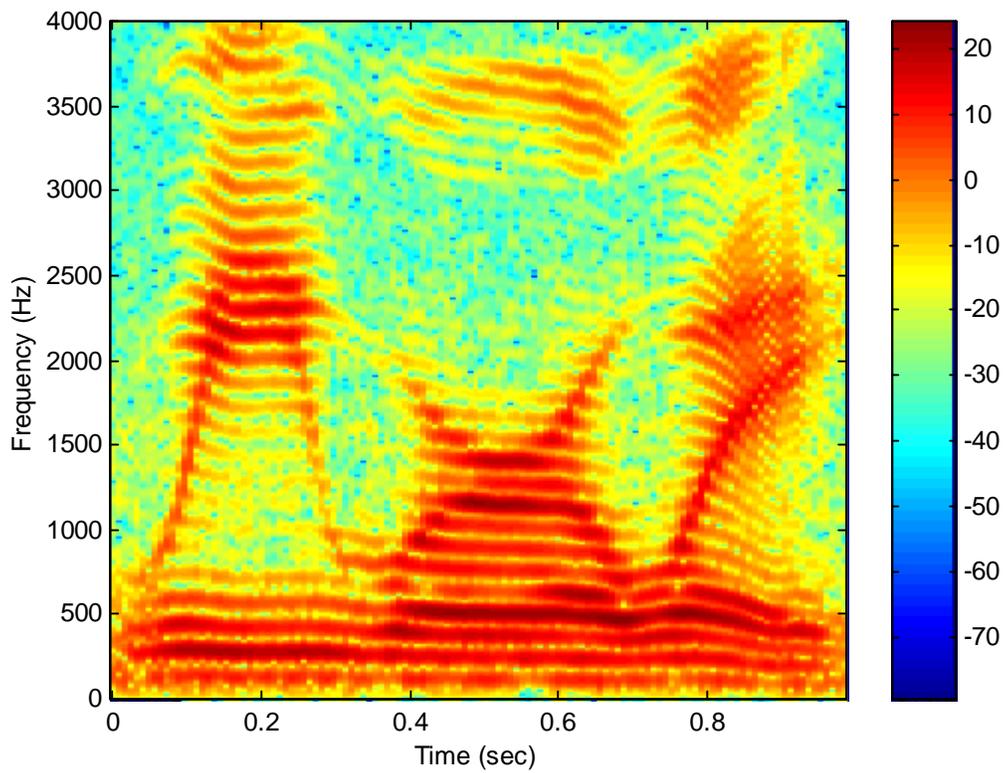

Figure 3.13- Narrowband Spectrogram for the phrase "*We were away*".

The estimation can be organized in two stages: (a) estimate the all-pole model H(z), (b) estimate the formants from H(z). The overall performance can be improved with sophisticated algorithms for the second stage. However, for comparison's sake, this work implemented a simple approach to extract the formants from H(z). All algorithms used the same function, so their performance on estimating H(z) can be compared. Given the estimated all-pole model $H(z) = \sigma / (1 - A(z))$, the first N formants were estimated as the first N roots of $1 - A(z)$ with frequency greater than 5 Hz. The implemented Matlab function *formants_fir* is shown in the Appendix. Table 3.1 shows the range of frequencies for the first 3 formants. This information was not incorporated into the function *formants_fir*. The overall performance of all algorithms would improve with the incorporation of this kind of information.

Table 3.1- Range for the first three formants [Rabiner, 78].

| Formant | Minimum (Hz) | Maximum (Hz) |
|---|---|---|
| F1 (1st) | 270 | 730 |
| F2 (2nd) | 840 | 2290 |
| F3 (3rd) | 1690 | 3010 |

The results presented in this work did not apply pre-emphasis to the speech signal. Besides, in order to make the plots more visually pleasant the formant estimation for the adaptive algorithms were plotted just every other 64 samples.

*Levinson-Durbin algorithm:*

For the Levinson-Durbin algorithm each frame was multiplied by a 20 ms Hamming window and a P-order A(z) was calculated through the *lpc* function [Snell, 93]. The window was shifted by 10 ms, (overlap of 50%). Figure 3.14 shows the first 3 formants when P = 8.

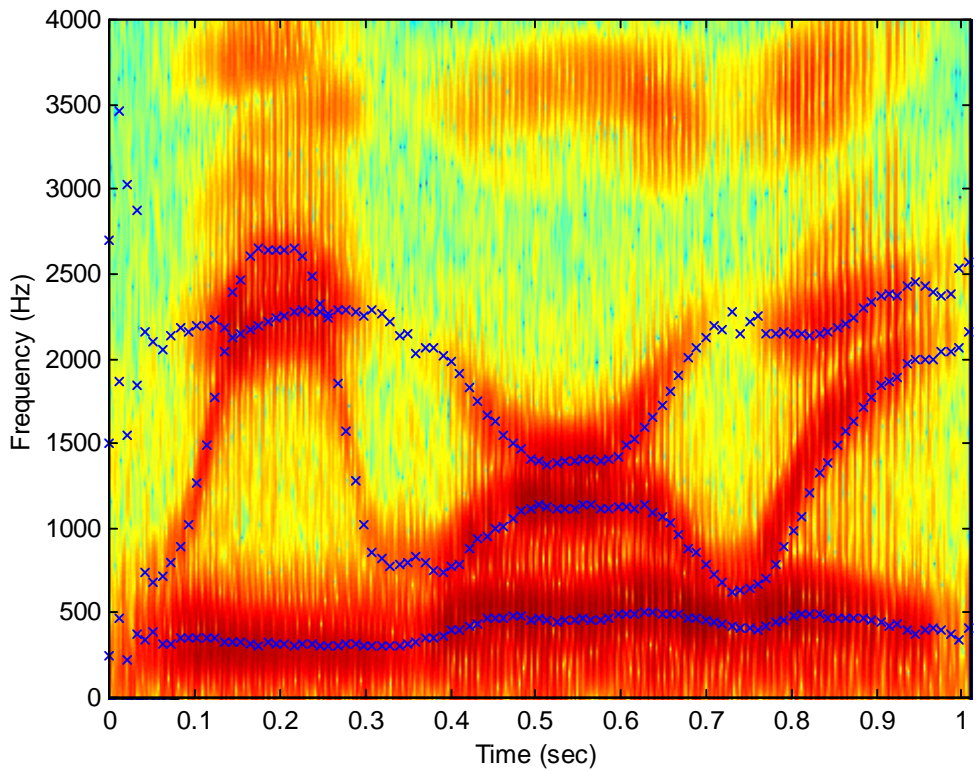

Figure 3.14- First three formants estimated through Levinson-Durbin algorithm, A(z) with P = 8 coefficients.

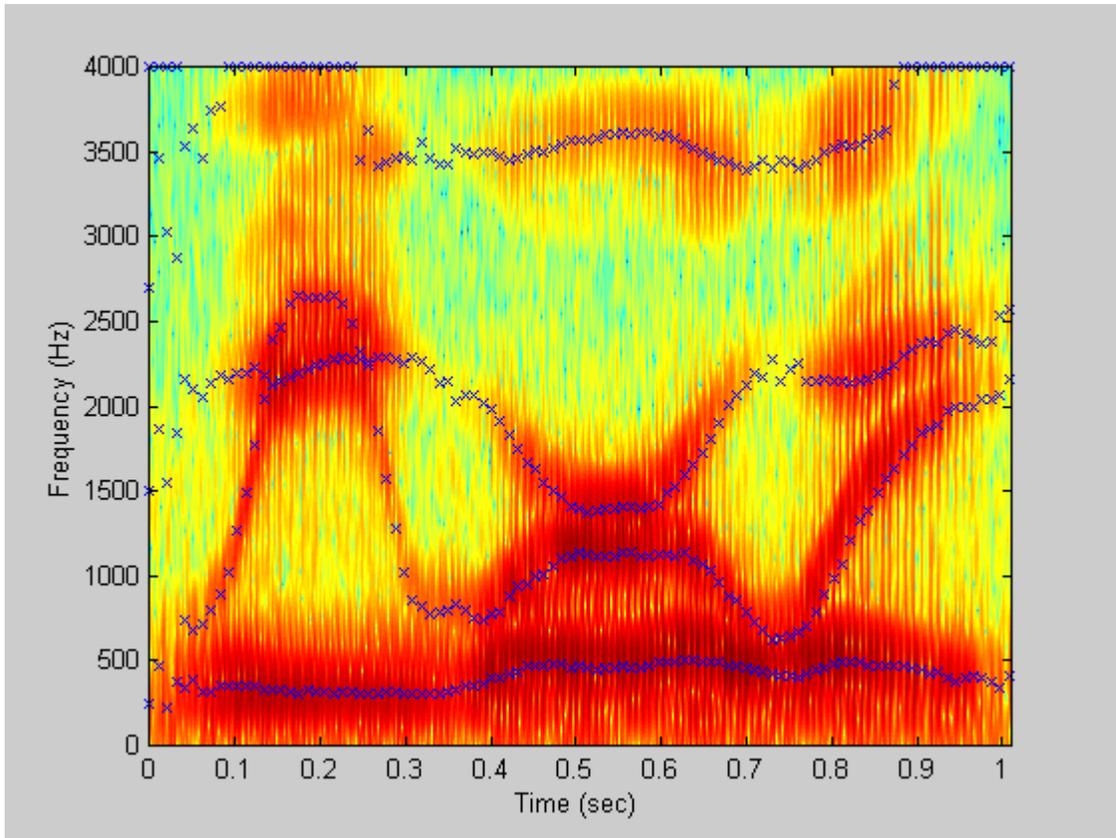

Figure 3.15- First four formants estimated through Levinson-Durbin algorithm, A(z) with P = 8 coefficients.

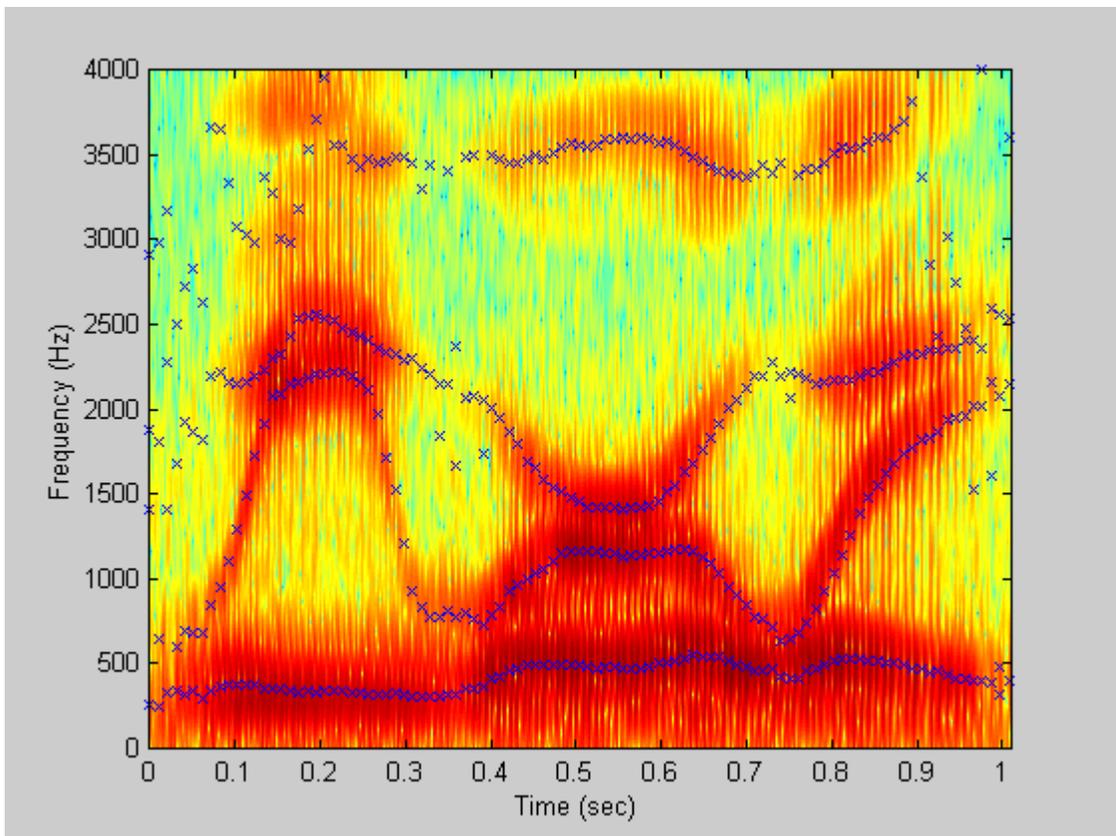

Figure 3.16- First four formants estimated through Levinson-Durbin algorithm, A(z) with P = 10 coefficients.

Several values of P were tested. Theoretically, for estimating N formants, P = 2 N is enough. However, the algorithm sometimes associates real roots for 1 − A(z) or makes confusion associating two pair of roots to the same formant. For N = 3 formants, the best results were obtained with P = 8, as shown in Figure 3.14. Figures 3.15 and 3.16 show the estimation for N = 4 formants, for P = 8 and 10, respectively. For N = 4, one can see that better results were obtained for P = 8 than P = 10. For P = 8, the values for the fourth formant F4 = 4,000 Hz (top of the plot) would be easily corrected with a post-processing based on Table 3.1. Figure 3.16 shows that between the words "we" and "were", when there is not much energy associated to F3 and F4, the algorithm tracked F4 but lost the F3 trajectory.

*LMS algorithm:*

Figures 3.17 and 3.18 show the performance of the LMS algorithm for estimating the first N = 3 formants with P = 8 for α = 0.1 and 0.2, respectively. One can see that when α = 0.1 the LMS was not able to track the fast decay in the trajectory of F2 in the end of the word "we". When α = 0.2 this problem is less accentuated, but the performance of the algorithm is not good in this region yet and the estimation is more noisy. Greater values of α led to a poor performance and the results in Figure 3.17 and 3.18 were considered the best results achieved with the LMS algorithm.

When one tries to estimate the first four formants (N = 4), the performance of the LMS algorithm is worse. Figure 3.19 shows the best result obtained when P = 8 and α = 0.2. Greater values for α result in a too noisy estimation and smaller values do not allow the algorithm to track the fast variations in the formant frequencies.

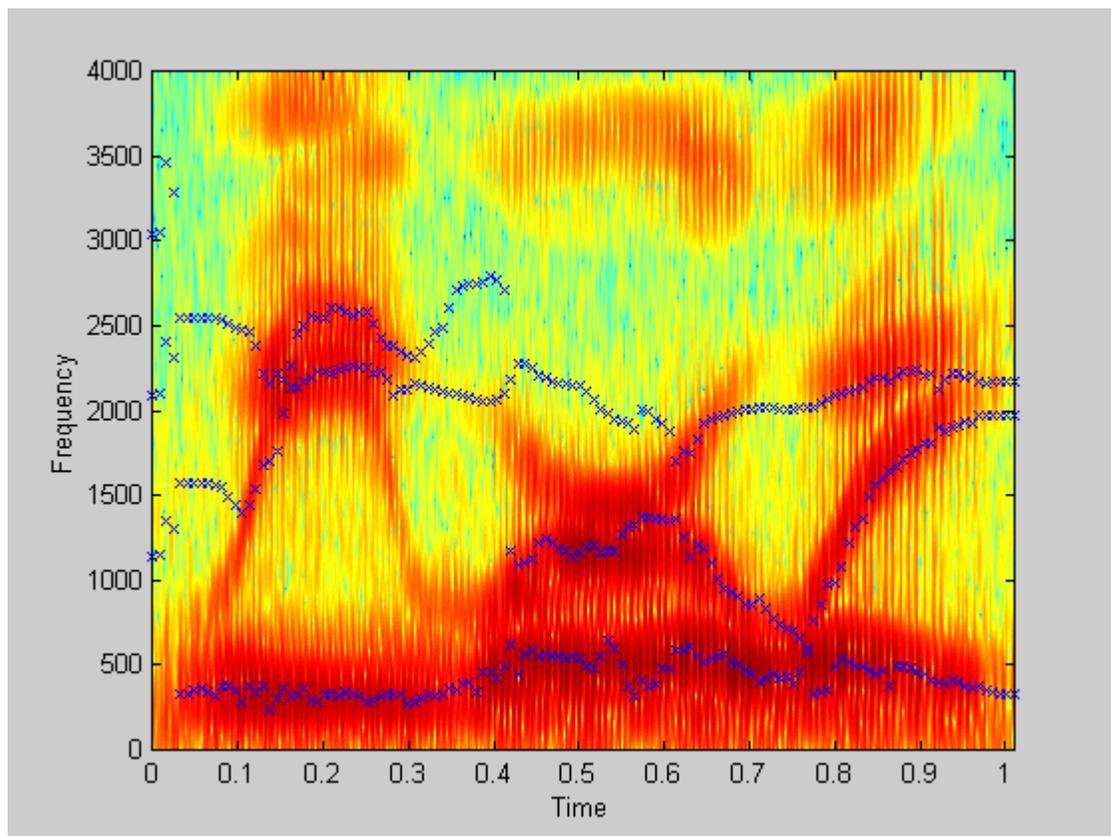

Figure 3.17- First three formants estimated through LMS algorithm, A(z) with P = 8 coefficients and α = 0.1.

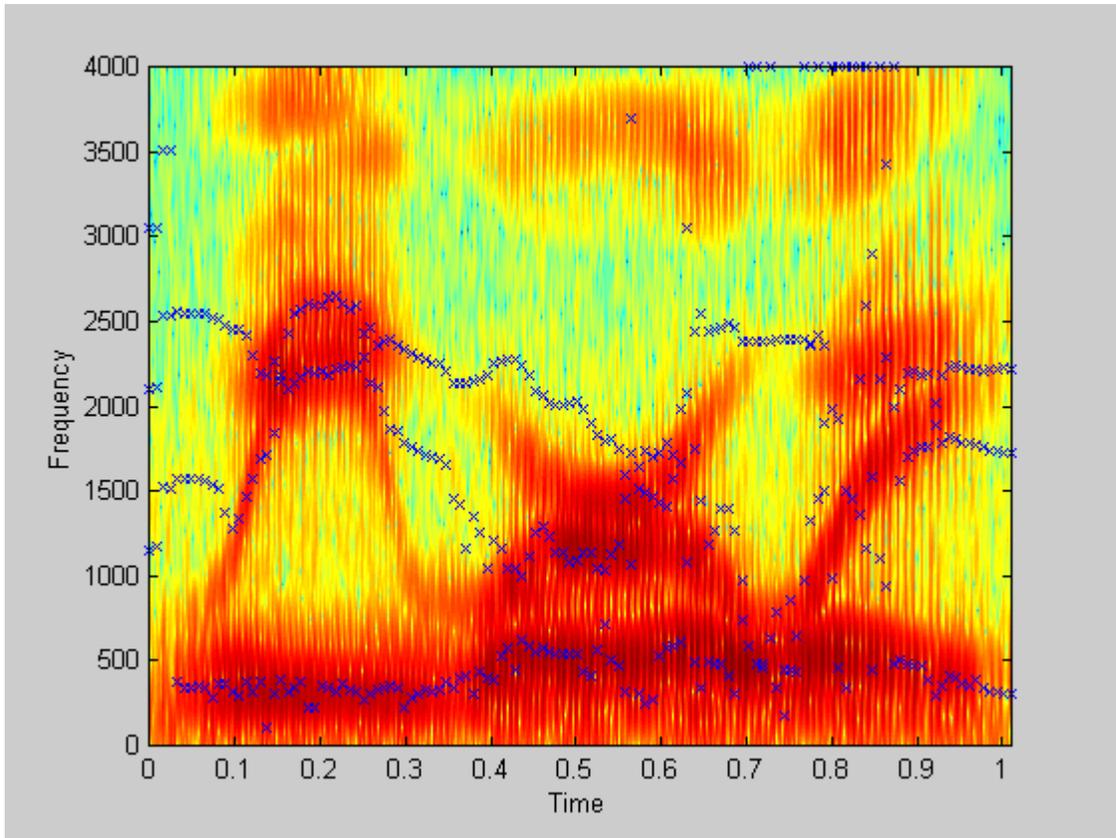

Figure 3.18- First three formants estimated through LMS algorithm, A(z) with P = 8 coefficients and α = 0.2.

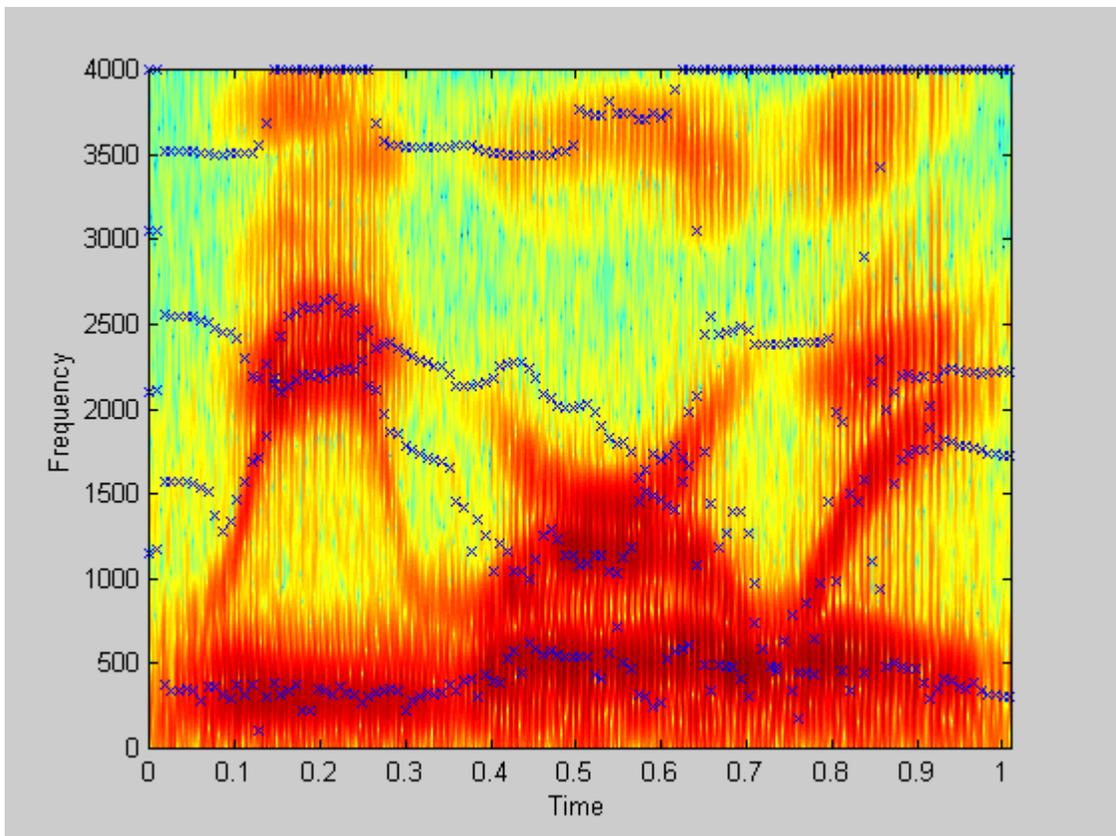

Figure 3.19- First four formants estimated through LMS algorithm, A(z) with P = 8 coefficients and α = 0.2.

*RLS algorithm:*

Figure 3.20 shows the performance of the RLS algorithm for estimating the first $N = 3$ formants with $P = 8$ and forgetting factor $\lambda = 0.8$. The RLS algorithm was able to track the formants in most regions as shown in Figure 3.20, but the estimation was too noisy.

Figure 3.21 shows the results when $\lambda = 0.99$ and $P = 8$ for estimating $N = 3$ formants. It can be seen that the RLS achieves a very good performance, similar to the one obtained for the Levinson-Durbin algorithm (Figure 3.14).

Figure 3.22 shows the results when $\lambda = 0.99$ and $P = 8$ for estimating $N = 4$ formants. It can be seen that the RLS achieves a very good performance again. The formants trajectories shown if Figure 3.22 can be even considered slightly smoother than the ones obtained through the Levinson-Durbin algorithm (Figure 3.15).

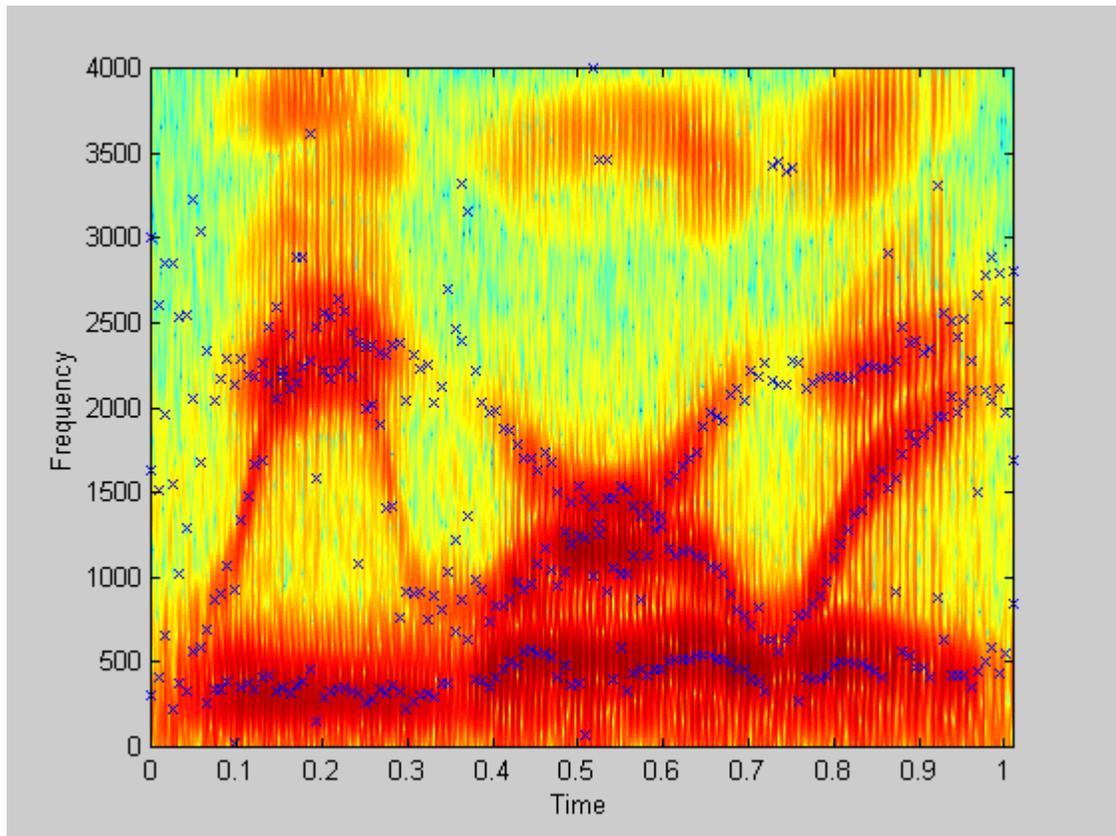

Figure 3.20- First three formants estimated through RLS algorithm, A(z) with P = 8 coefficients and $\lambda = 0.8$.

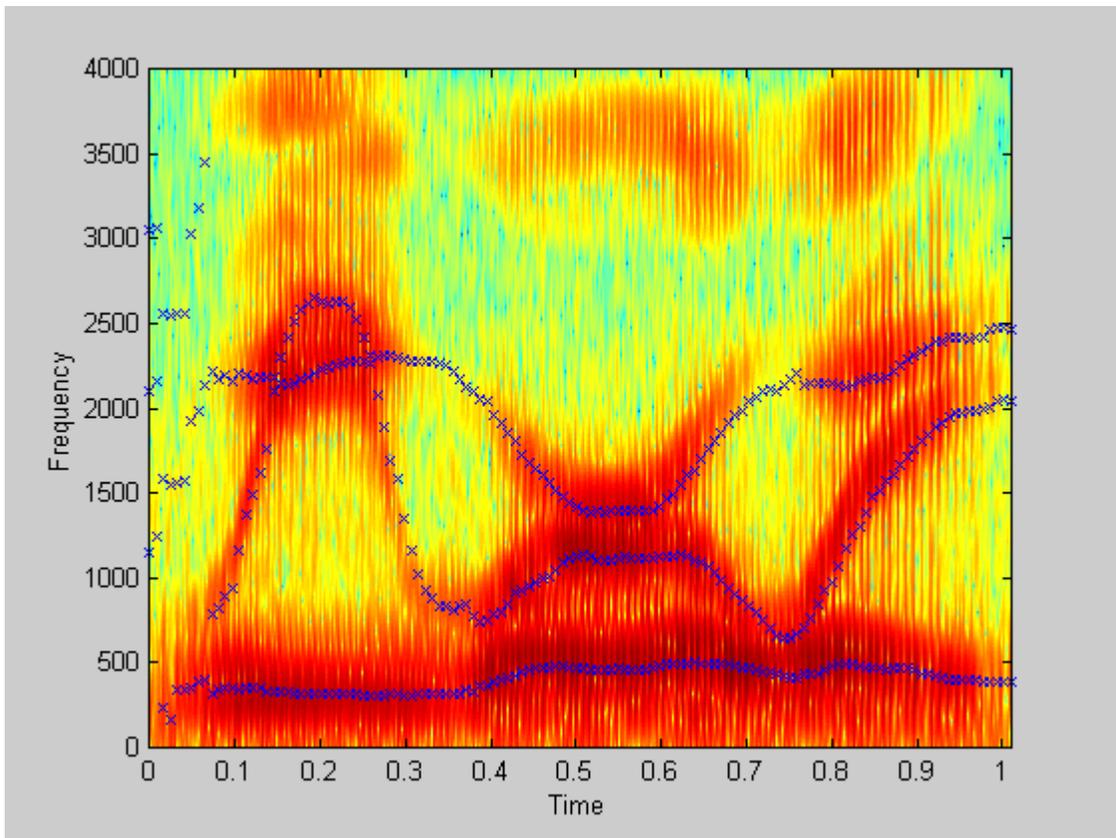

Figure 3.21- First three formants estimated through RLS algorithm, A(z) with P = 8 coefficients and λ = 0.99.

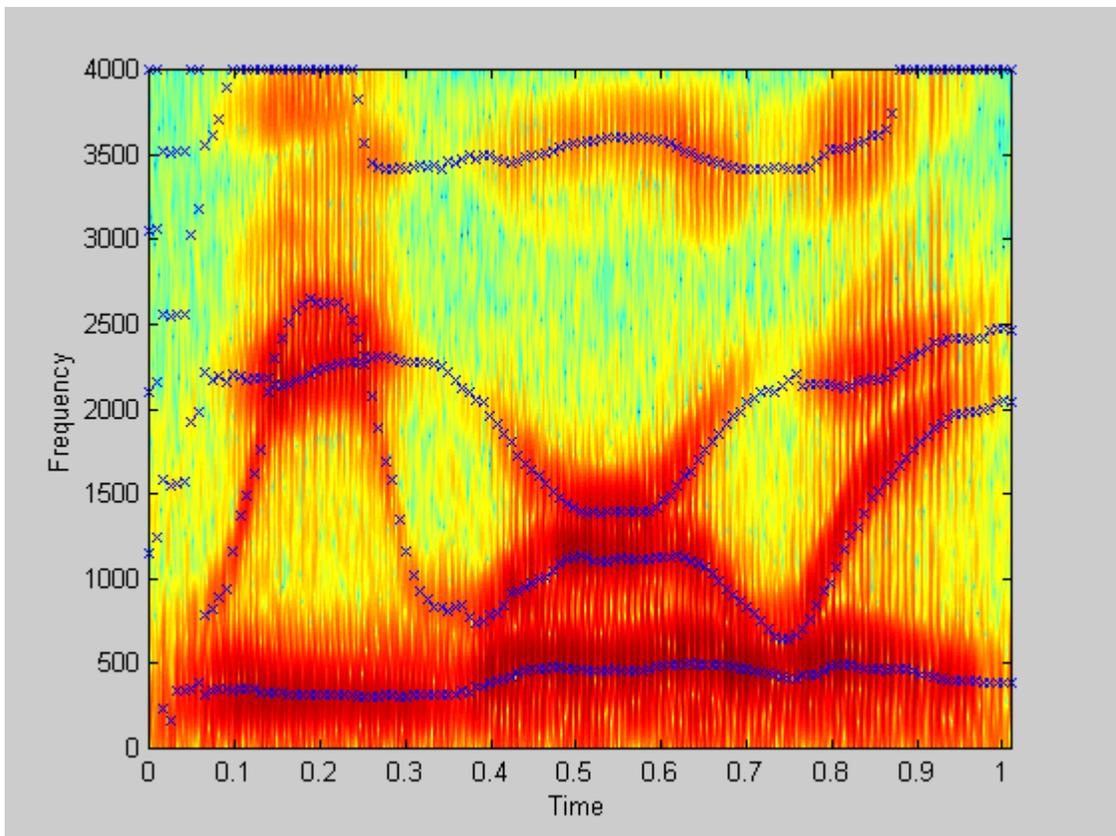

Figure 3.22- First four formants estimated through RLS algorithm, A(z) with P = 8 coefficients and λ = 0.99.

*Computational Complexity:*

The previous results showed that the RLS algorithm achieves a much better performance than the LMS algorithm in the task of estimating formants. However, as pointed out in section 2, this improvement in performance is achieved at the expense of a large increase in computational complexity. Thus, because complexity is an important issue in this case, I intended to estimate it.

The Matlab allows a first-order estimation of the number of floating-point operations through its function *flops*. Table 3.2 shows a comparison among the algorithms used in this work. The complexity of the LMS algorithm in the task of estimating the formants of "We were away", that corresponds to 285,075 "flops", serves as basis for comparison. It can be seen that the RLS takes approximately 52 times more operations than the LMS, and more than 2 times more operations than Levinson-Durbin.

Table 3.2- Comparison of computational complexity.

| Algorithm | LMS | Levinson-Durbin | RLS |
|---|---|---|---|
| Complexity | 1 | 21 | 52 |

# 4) Conclusions

This work showed a comparison between the LMS and RLS algorithms for the task of estimating the speech formant frequencies. The results obtained by these adaptive algorithms were compared to the results obtained by the block-based Levinson-Durbin algorithm.

It was showed that the conventional LMS algorithms does not have a good performance in this task, due mainly to the highly time-variant environment and the eigenvalues spread. The first formant F1, that is usually associated to a greater energy than the other formants, can be tracked with reasonable efficiency by the LMS. However, tracking simultaneously more than one formants is a task that seems to be too difficult to the conventional LMS. The results showed that even with a careful tuning of the parameters $\alpha$ and P, the LMS algorithms did not track the first four formants.

The RLS algorithm achieved an excellent result, that is comparable to the results obtained by the Levinson-Durbin algorithm. However, the computational complexity of RLS is 2 times greater than the Levinson-Durbin's. Both algorithms have a computational complexity much greater than the LMS algorithm. The choice among these three options can be considered as application dependent.

In applications that are inherently block-based, as modern speech coding, the Levinson-Durbin is a natural choice. On the other hand, applications in the area of speech recognition can exploit the advantage of the sample-basis approach of the adaptive algorithms. For example, there are interest on designing computer games as tools for training deaf children to speak correctly [Javkin, 93], [Javkin, 94], [Kim, 94], [Levitt, 73]. In this context, the RLS algorithm seems to be more appropriate than the Levinson-Durbin algorithm. The designer can tune the forgetting parameter $\lambda$ to construct an application with output (visual feedback for the children) in a sample-by-sample basis and with smooth formant trajectories.